\documentclass[aps,prb,twocolumn,superscriptaddress,longbibliography]{revtex4-2}

\usepackage{graphicx,bm,amssymb,amsmath,xcolor,pifont,soul,multirow}
\usepackage[linktocpage=true,colorlinks=true,pdfborder={0 0 0},linkcolor=blue,citecolor=blue,filecolor=yellow,urlcolor=blue,bookmarks,pdfauthor={},]{hyperref}
\usepackage[normalem]{ulem}
\usepackage{epstopdf}
\usepackage{epsfig}
\usepackage{blkarray}
\usepackage{makecell}

\def\bk{{\bf k}}

\def\>{\rangle}
\def\<{\langle}
\usepackage{amsmath}

\setstcolor{red}

\definecolor{dgreen}{rgb}{0.0, 0.8, 0.6}

\usepackage{titlesec}
\usepackage{lipsum}

\titleformat{\subsubsection}
  {\normalfont}{\thesubsubsection}{1em.}{}
\renewcommand{\thesubsubsection}{\roman{subsubsection}}

\begin{document}
\title{Improving structure search with hyperspatial optimization and TETRIS seeding}

\author{Daviti Gochitashvili}
\affiliation{Department of Physics, Applied Physics, and Astronomy, Binghamton University-SUNY, Binghamton, New York 13902, USA}
\author{Maxwell Meyers}
\affiliation{Department of Physics, Applied Physics, and Astronomy, Binghamton University-SUNY, Binghamton, New York 13902, USA}
\author{Cindy Wang}
\affiliation{Barrington High School, Barrington, Illinois, 60010, USA}
\author{Aleksey N. Kolmogorov}
\affiliation{Department of Physics, Applied Physics, and Astronomy, Binghamton University-SUNY, Binghamton, New York 13902, USA}
\date{\today}

\begin{abstract}
Advanced structure prediction methods developed over the past decades include an unorthodox strategy of allowing atoms to displace into extra dimensions. A recently implemented global optimization of structures from hyperspace (GOSH) has shown promise in accelerating the identification of global minima on potential energy surfaces defined by simple interatomic models. In this study, we extend the GOSH formalism to more accurate Behler-Parrinello neural network (NN) potentials, make it compatible with efficient local minimization algorithms, and test its performance on nanoparticles and crystalline solids. For clusters modeled with NN potentials, four-dimensional optimization offers fairly modest improvement in navigating geometric relaxation pathways and incurs increased computational cost largely offsetting the benefit, but it provides a significant advantage in facilitating atom swaps in nanoalloys. In comparison, the introduction of a moderate, controlled bias for generating more physically sensible starting configurations, achieved via TETRIS-inspired packing of atomic blocks, has a more direct impact on the efficiency of global structure searches. The benchmarked systems are Lennard-Jones clusters, Au or Cu-Pd-Ag nanoparticles and binary Sn alloys described by NN potentials, and compounds with covalent B or BC frameworks modeled with density functional theory.
\end{abstract}	

\maketitle

\section{Introduction}
\label{sec:introduction}

Accurate and efficient prediction of atomic structures has evolved from an intractable challenge in the 1990s to an integral component of the materials discovery process ~\cite{Woodley2008,Wales1999,Zurek2016,Oganov2006}. Advanced structure search methods are now routinely used to identify crystalline and nanoscale configurations synthesizable under specific experimental conditions~\cite{Needs2016,Oganov2019,Woodley2020}. Prediction of thermodynamically stable compounds with unique morphologies beyond known prototypes~\cite{matpro,aflow,oqmd,icsd} has been particularly important in the exploration of matter subjected to extreme pressures. Notable predictions that guided materials synthesis or helped interpret experimental results include the ground states of elemental Na~\cite{Dong2017} and B~\cite{Oganov2009} exhibiting unusual properties under compression; LiB~\cite{ak08,ak30}, MB$_4$ (M = Cr, Mn, or Fe)~\cite{ak16,ak17,ak28}, and CaB$_6$~\cite{ak23,ak24} metal borides featuring unfamiliar frameworks; Na$_2$IrO$_3$~\cite{ak21}, Na$_3$Ir$_3$O$_8$~\cite{ak36}, and Cu$_2$IrO$_3$~\cite{ak44} iridates showcasing rich strongly correlated physics; and SiH$_4$~\cite{Pickard2009,Pickard2022}, H$_3$S~\cite{Li2016}, and other compressed hydrides boasting record superconducting critical temperatures.

The complex optimization problem of locating global minima on multidimensional potential energy surfaces in given chemical spaces has been addressed with a variety of strategies implemented in USPEX~\cite{Oganov2006}, XtalOpt~\cite{LONIE2011}, CALYPSO~\cite{WANG20122063}, AIRSS~\cite{Pickard2011}, MAISE~\cite{ak41}, and other packages. The methodological diversity reflects the implications of the no-free-lunch theorem, which posits that no single optimization algorithm can outperform all others across every possible system~\cite{Wolpert1997}. While the developed algorithms are often fine-tuned for specific materials classes, such as metal alloys~\cite{Wolverton1998, Mueller2009}, molecular crystals~\cite{Desiraju2013,Hunnisett2024}, or nanoparticles~\cite{Johnston2003,Ferrando2008}, their overall efficiency is largely determined by common factors: how structures are generated, evolved, and selected. 

Creation of diverse yet physically meaningful starting configurations is critical for a large-scale efficient screening and ranges from purely random to highly constrained. It has been demonstrated that the introduction of atom exclusion zones~\cite{Woodley2004a,Woodley2004b}, incorporation of lattice parameters from experiment~\cite{ak23}, seeding searches with supercells of randomized known ground states~\cite{ak23, ak36}, or generation of crystal structures with randomly chosen symmetries ~\cite{Avery2017, Fredericks2021} can speed up the determination of ground states by orders of magnitude. 

Equally important is to control bias in structure optimization. Local relaxation of starting configurations with deterministic gradient-based algorithms introduces minimal bias and has been used extensively in {\it ab initio} random structure searches (AIRSS)~\cite{Pickard2011}. Variable degrees of implicit steering toward global minima can be achieved with the minima hopping method~\cite{Wales1997}, particle swarm optimization~\cite{Wang2010}, evolutionary algorithm~\cite{DAVEN1996,Woodley2004,Oganov2006}, and other strategies that allow structure populations to escape from local minima. Modified variable-composition~\cite{Trimarchi2009,Zhu2014,Lepeshkin2019} and variable-size multitribe~\cite{ak38} evolutionary algorithms further facilitate the sharing of favorable motifs identified in large configuration spaces to iteratively construct improved configurations. The introduction of fingerprint-based energy penalty for low-symmetry structures has been shown to enhance convergence in various systems~\cite{Tao2024}.

Screening candidate structures with classical interatomic potentials can be a promising means of accelerating {\it ab initio} structure searches. Inexpensive traditional interaction-specific models have been used extensively in studies of nanoclusters~\cite{Ferrando2008} but their generally limited transferability can lead to finding suboptimal solutions~\cite{ak38,ak40}. Recently developed machine learning interatomic potentials offer significantly higher accuracy~\cite{Behler2016, Bartok2017} and have been successfully used to identify more favorable nanocluster configurations~\cite{ak38,ak40} and thermodynamically stable compounds~\cite{ak47,ak51} overlooked in previous {\it ab initio} searches.

An out-of-the box idea explored decades ago is based on performing global searches in higher dimensions. It was proposed that introducing extra spatial dimensions could help structures escape local minima by enabling exploration of directions unavailable in 3D space~\cite{Faken1999,Bayden2004}. In these studies, structures were first minimized in 3D, then displaced into a higher-dimensional configuration space where they performed constant-energy walks, before being returned to 3D for further relaxation. The added degrees of freedom provided access to relaxation pathways that more effectively traverse barriers on the potential energy surface, increasing the likelihood of reaching low-energy configurations. This strategy proved particularly useful for systems with complex landscapes, such as the double-funnel topology of the Lennard-Jones (LJ) 38-atom cluster. The unorthodox approach was investigated further in a recent study~\cite{Pickard2019}. The global optimization of structures from hyperspace (GOSH) generates atomic configurations stochastically in higher special dimensions and gradually brings them back to the real 3D space during a local optimization. The algorithm was found to outperform the traditional 3D local relaxation for several non-periodic systems, from unary and binary clusters to linear chains and covalently bonded frameworks, modeled with simple pairwise potentials. The probability enhancement of finding the ground state reached two orders of magnitude in some cases and was especially evident for the connected systems~\cite{Pickard2019}. One of the open questions was how effective the hyperdimensional optimization would be in the exploration of energy surfaces defined by more sophisticated potentials.

The present study focuses on implementing and examining two structure generation and optimization factors. First, we investigate the performance of GOSH by extending the formalism to neural network potentials (NNPs). We have modified our MAISE package~\cite{ak41} to perform GOSH for clusters and crystals described with more complex interatomic NNPs of the Behler-Parrinello type~\cite{Behler2007} that provide near-{\it ab initio} accuracy. Our implementation reproduces and enhances the previously observed efficiency gains for LJ monoatomic clusters but shows little to no advantage over the deterministic 3D optimization in the case of Au clusters or Sn crystalline alloys, especially when the increased computational cost is taken into account. On the other hand, a notable improvement is observed for multicomponent clusters.

Second, we generalize our previously proposed TETRIS generation of atomic configurations~\cite{ak38} to periodic materials and benchmark its performance against more stochastic population initialization. The scheme was originally designed for creating compact starting nanoparticle configurations with physically meaningful interatomic distances by shooting atoms toward the cluster’s center iteratively and keeping the closest positions. Our present tests reveal that this essentially parameter-free approach offers a better efficiency improvement than GOSH does for monoatomic LJ and multicomponent Cu-Pd-Ag clusters. The implementation of the TETRIS scheme for crystalline compounds enables a controlled injection of bias in the form of atomic blocks and targeted number of bonds. We demonstrate that the method has little effect on the probability of finding ground states in M-Sn alloys (M = Li, Pd, or Ag) with predominantly metallic bonding~\cite{ak47,ak51} but offers significant improvement in locating more stable configurations in a layered metal borocarbide (ambient-pressure LiB$_2$C$_2$)~\cite{ak52} and a metal boride (CaB$_6$ at 50 GPa)~\cite{ak23} featuring complex covalent networks. The customizable TETRIS seeding protocol, made available in the open-source MAISE package~\cite{ak41}, holds promise for accelerating the convergence of global optimization searches via a judicious introduction of beneficial motifs in structure populations.

\section{Methods}
\label{sec:methods}

\subsection{ GOSH original formulation} 

The global optimization method designed by Pickard~\cite{Pickard2019} relies on generating atomic arrangements in higher dimensions, following a downhill trajectory on a hypersurface, and eventually forcing the system to materialize in the normal space. The creation of non-periodic structures involves random placement of hard hyperspheres of radius $\tilde{a}_0$ within a confining hypersphere of volume $\tilde{V}/f$, where $\tilde{a}_0$ is half the expected equilibrium bond length, $\tilde{V}$ is the total volume of atom centered hyperspheres, and $f$ is the packing fraction. The potential energy surface in the hyperspace of dimension $d=d_0+d_+$ is a combination of the pairwise potential energy $V_{i,j}(\tilde{r}_{i,j})$ and a harmonic penalty term
\begin{equation}
\bar{E}(\{\tilde{\mathbf{x}}_i\}) = \tilde{E}(\{\tilde{\mathbf{x}}_i\}) + \frac{1}{2}\mu \sum_{i} l_i^2 
= \sum_{i,j} V_{ij}(\tilde{r}_{ij}) + \frac{1}{2}\mu \sum_{i} l_i^2,
\nonumber
\end{equation}
where the Euclidian distances are defined as
\begin{align}
\tilde{r}_{i,j} &= \left[\sum_{n=1}^d (\tilde{x}_{j,n} - \tilde{x}_{i,n})^2\right]^{1/2}, \\
l_i &= \left[\sum_{n=d_0+1}^d (\tilde{x}_{i,n})^2\right]^{1/2}. \nonumber
\end{align}
The chosen two-step gradient descent algorithm that does not require line minimization allows for a continuous adjustment of the spring constant $\mu_{k+1}=\mu_k\beta$ after each atomic position optimization cycle $k$. The $\mu_0=10$ and $\beta=0.001$ values provided an efficient convergence of the hyperspacial energy gradients and ensured the fall of extraspatial coordinates below an acceptable threshold.

\subsection{ GOSH implementation in MAISE} As discussed by Pickard~\cite{Pickard2019}, GOSH can be naturally generalized for non-pairwise potentials and coupled with other minimization algorithms. In order to make the method compatible with the existing structure generation and optimization capabilities available in MAISE, we introduced a few modifications in the original scheme. 

\begin{figure}[t!]
   \centering
\includegraphics[width=0.48\textwidth]{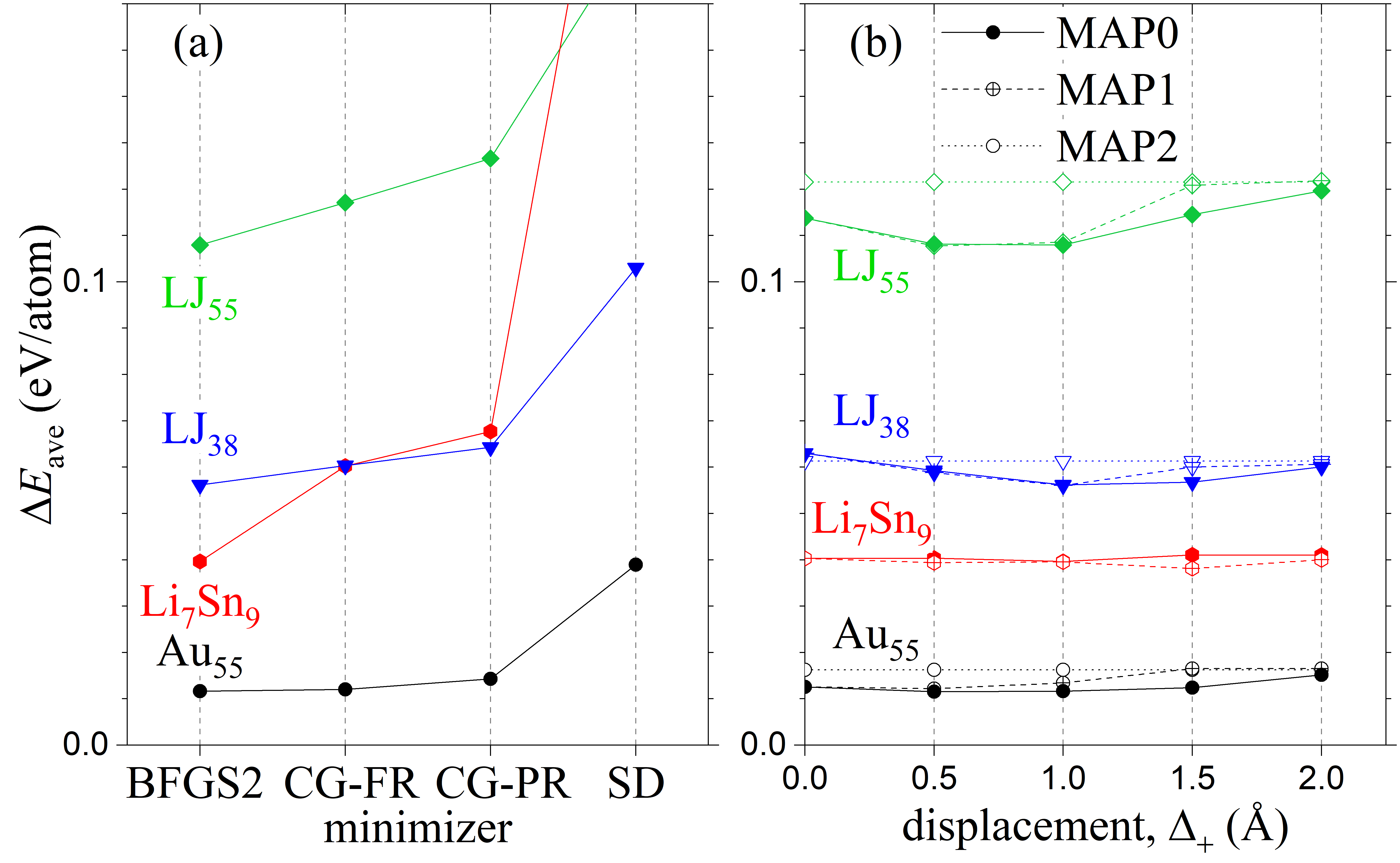}
    \caption{\label{fig-01} Performance of different protocols in hyperdimensional optimization of LJ or Au clusters and Li-Sn crystals. $\Delta E_{\textrm {ave}}$ represents the average energy of the best half of 2,000 structures relative to the corresponding ground state. (a) The local minimization algorithms were tested for MAP0 and $\Delta_+=1.0$ \AA. (b) The mapping and displacement variants were tested with BFGS2. The MAP2 scheme is independent of $\Delta_+$.}
\end{figure}

First, we enabled access to a wider range of local optimization algorithms with line minimization available in the GSL~\cite{galassi2018gnu} by switching from continuous to stepwise change in the spring constant $\mu$. We break down the maximum number of minimization steps $M$ into $N$ equal groups and set $\mu_n=\mu_0 \beta^n$. In the last set, we zero the extraspatial coordinates, which are almost always negligible at this stage, to ensure that the structure is fully optimized in the normal space. We observed little dependence of the optimization efficiency on these settings and used $M=600$, $N=6$, $\mu_0=2$, and $\beta=2$ (for ease of comparison, we keep specifying unitless $\mu$ values but scale the parameter as eV/\AA$^2$ and set the pairwise LJ energy parameter to 1 eV in our simulations). Based on the benchmarking tests for Broyden–Fletcher–Goldfarb–Shanno (BFGS2), Fletcher-Reeves conjugate gradient (CG-FR), Polak-Ribiere conjugate gradient (CG-PR), and steepest descent (SD) algorithms~\cite{fletcher1964,fletcher1970,Klessig1972,Fliege2000} in Fig. \ref{fig-01} (a), we chose BFGS2 as the default minimizer.

\begin{figure}[t!]
   \centering
\includegraphics[width=0.48\textwidth]{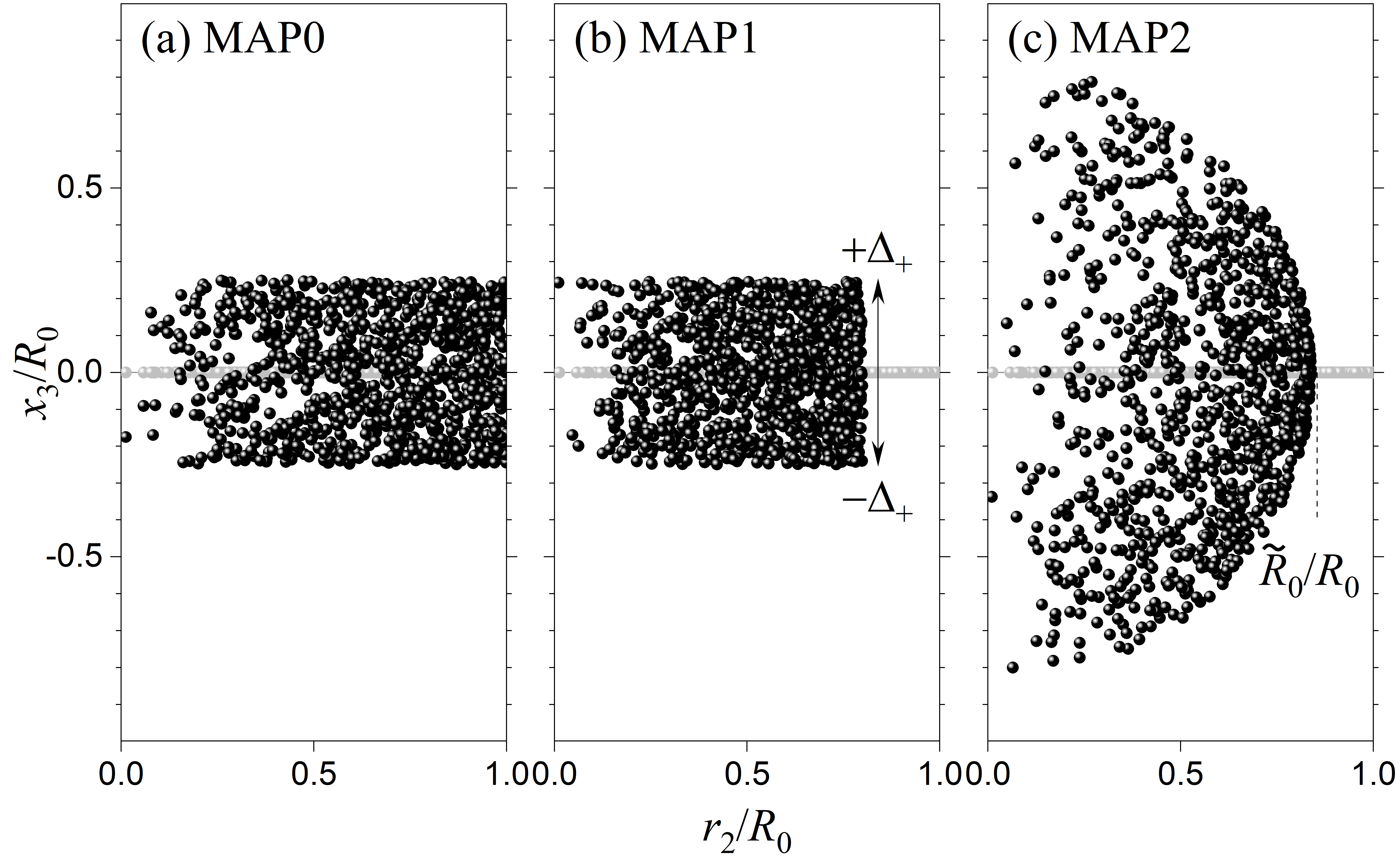}
    \caption{\label{fig-02} Atom distributions in 20 50-atom 2D clusters (in gray) expanded into 3D (in black) using different mapping protocols. The hyperdimensional $x_3$ coordinate is plotted as a function of the 2D radial distance normalized to the largest radius $R_0$. (a) $x_3$ is generated randomly in $[-\Delta_+;\Delta_+]$. (b) After the random $x_3$ generation, the 2D coordinates are contracted. (c) The 2D coordinates are remapped and $x_3$ is generated randomly in the $\pm\sqrt{\tilde{R}_0^2-r^2_i}$ range as detailed in the main text.}
\end{figure}

Second, we implemented alternative procedures for extending structures beyond the normal space. For clusters, we aimed to retain the possibility of initializing configurations with the efficient TETRIS algorithm or generating offspring with various mutation or crossover operations in evolutionary searches~\cite{ak38,ak40,ak41}. Therefore, instead of creating clusters with equally randomized atomic coordinates in all dimensions, we enabled random shifts into the extra dimensions for clusters with already defined 3D atomic positions. Protocol MAP0 involves uncorrelated random displacements of atoms into each extra dimension up to $\Delta_+$ (for $d_0=2$ and $d_+=1$, this operation transforms a 2D disk into a fuzzy flat cylinder). Optionally, the normal coordinates are uniformly contracted to compensate for the increase of the interatomic distances in the hyper space (MAP1). Fig.~\ref{fig-01}(b) illustrates a good performance of the modified GOSH for $\Delta_+\sim 1$ \AA. The last tested protocol, MAP2, treats the normal and extra dimensions on a more equal footing and maintains a roughly uniform atomic density via a correlated adjustment of the normal and hyper coordinates (for $d_0=2$ and $d_+=1$, it converts a 2D disk into a near-spherical 3D shape). If the maximum distance to the origin in a cluster with $N$ atoms is $R_0$ in the normal space and becomes $\tilde{R}_0$ upon expansion into one extra dimension, the normal space radial distance $r_{d_0}^{\rm old}$ for each atom can be mapped approximately as 
\begin{equation}
r_{d_0} = \tilde{R}_0 \left[ 1- \left( 1 - \left(\frac{r^{\textrm{old}}_{d_0}} {R_0}\right)^2\right)^{p} \right]^{1/2},\nonumber
\end{equation}
with $p=0.625$, $2/3$, and $0.694$ for 1D, 2D, and 3D, respectively (see Supplemental Note I for further detail~\cite{SM}). The extra dimensional coordinate $x_{d_0+1}$ should be generated randomly in the $\pm\sqrt{\tilde{R}_0^2-r^2_{d_0}}$ range. The $\tilde{R}_0/R_0$ ratios, assessed by rearranging $N$ atoms on uniform grids in $d_0$ and $d_0+1$ dimensions, are $1.110\;N^{-1/6}$ for $d_0=2$ and $1.082\;N^{-1/12}$ for $d_0=3$. Since Fig.~\ref{fig-01}(b) illustrates a generally good performance of the MAP0 protocol with $\Delta_+=1$ \AA, we have used it in all present simulations unless specified otherwise.

For crystals, the normal and extra dimensions are inherently different because the periodicity over hyperspatial coordinates, once introduced, cannot be naturally phased out. Therefore, we chose to extend crystalline structures with 3D periodicity by introducing random hyperspatial positions up to $\pm \Delta_+=\pm 1$ \AA\ and uniformly shrinking the 3D volume by 0.8. It should be noted that the normal space contraction compensating for the average bond elongation due to the random shifts in the extra dimensions may bring some atoms too close to each other. To avoid that, we perform a series of hyperspatial coordinate adjustments with a simple repulsive potential for both clusters and crystals until all interatomic distances are larger than $0.7$ of the expected equilibrium length, just as done for making random 3D structures~\cite{ak38,ak41}.

\subsection{ TETRIS algorithm}

MAISE features several options for generating initial structures~\cite{ak41}. A standard BLOB approach for constructing clusters relies on randomly placing atoms around the origin and tweaking their positions to ensure realistic interatomic distances with a combination of a repulsive pairwise potential and attractive springs that keep atoms within a targeted cluster radius. An alternative TETRIS-inspired algorithm involves sending atoms from random directions toward the origin one by one and keeping the closest position out of $N^{3/2}$ tries to create spherical or ellipsoidal shapes~\cite{ak38}. Our previous studies illustrated benefits of initializing evolutionary searches with TETRIS-generated configurations. Supplementary Fig. 3~\cite{SM} confirms that TETRIS clusters are more compact than BLOB ones, with the volume definition adopted from Ref.~\cite{Chiriki2017}, even after the latter procedure is tightened with a compression factor of 1.3.

We now extend the method to crystals and introduce an option to seed structures with predefined motifs. While packing algorithms are popular for determining molecular crystal ground states~\cite{Desiraju2013, Hunnisett2024}, the use of building blocks is uncommon in explorations of inorganic solid-state materials. Our TETRIS-based generation proceeds as follows: (i) create random lattice vectors; (ii) extend the $c$ lattice vector to temporarily disable periodicity along this direction; (iii) randomly select blocks, producing the desired composition, from a user-defined set; (iv) optionally perform a random 3D rotation of each block using quaternions; (v) send blocks one by one from a random lateral position along the $c$ axis, trying all possible orientations in the $x$-$y$ plane at each downward step, until they either reach the bottom or come into close contact with another block; (vi) repeat the previous step for several starting lateral positions and keep the position with the lowest $z$; and (vii) once all the blocks are dropped, shrink $c$ until atom separations match the shortest allowed distance. 

Due to the observed insignificant influence of implementation details on the method’s overall performance, the degree of introduced bias can be controlled by customizing the set of atomic blocks. Additionally, one can specify the targeted number of nearest neighbors within a cutoff sphere for each species to discard configurations with unphysical local environments. Our tests illustrate that the designed algorithm works particularly well for generating layered covalent frameworks.

\subsection{ DFT and NN calculations}

For metal borides and borocarbides, DFT calculations were performed with {\small VASP}~\cite{Kresse1996} using projector augmented wave potentials~\cite{Blochl1994} and the optB88-vdW functional~\cite{optB88}. A 500~eV plane-wave cutoff and dense Monkhorst-Pack $\bk$-meshes~\cite{Monkhorst1976} with $\Delta k \sim 2 \pi \times 0.025$~\AA$^{-1}$ ensured numerical convergence to typically within 1 meV/atom.

Detailed information about our previously developed NN interatomic potentials for Au, Cu-Pd-Sn, and Sn alloys, available on GitHub, can be found in Refs.~\cite{ak38,ak40,ak41,ak47,ak51}. The local atomic environments were represented with 51 Behler-Parrinello symmetry functions per element, with the explicit expressions provided in Supplemental Note II~\cite{SM} and the parameters detailed in Ref.~\cite{ak38}. The details of NN models of different systems studied in this paper is summarized in Table~\ref{tab:test-errors}. NNs were fitted to PBE-level energies and forces and extensively tested in global structure searches. The iterative generation of reference structure sets and NN training were done with the automated \small{MAISE-NET} wrapper~\cite{ak41}.

\begin{table}[h]
\centering
\begin{tabular}{lcccccc}
\hline
Chemical & \# of & \# of & \# of & testing & testing & reference\\ 
 system & weights & $E$ data & $F$ data & error $E$ & error $F$ & study \\
\hline\hline
Au      & 641 & 2,912  & 21,621 & 6.5 & 37 & \cite{ak40}\\

Cu-Pd      & 3,162 & 3,725  & 32,223 & 6.6 & 63 & \cite{ak38}\\
Cu-Ag      & 3,162 & 3,724  & 32,034 & 3.5 & 38 & \cite{ak38}\\
Pd-Ag      & 3,162 & 3,705  & 32,166 & 4.8 & 58 & \cite{ak38}\\
Cu-Pd-Ag   & 8,853 & 2,191  & 29,163 & 5.2 & 53 & \cite{ak38}\\

Li-Sn  & 3,162 & 6,046  & 46,410 & 10.2 & 49 & \cite{ak47}\\
Pd-Sn  & 3,162 & 5,507  & 46,272 & 9.6 & 52 & \cite{ak51} \\
Ag-Sn  & 3,162 & 5,410  & 37,548 & 8.4 & 33 & \cite{ak51}\\

\hline\hline
\end{tabular}
\caption{Neural network (NN) parameterization details by chemical composition, including the total number of adjustable weights, the number of energy and force data points in the training sets, and the root mean square errors on the test sets for energies (meV/atom) and forces (meV/\AA). All models were trained and evaluated using PBE-level~\cite{PBE} reference data.}
\label{tab:test-errors}
\end{table}

\section{Results and Discussion}
\label{sec:results}

\begin{figure}[t!]
   \centering
\includegraphics[height=0.3\textwidth]{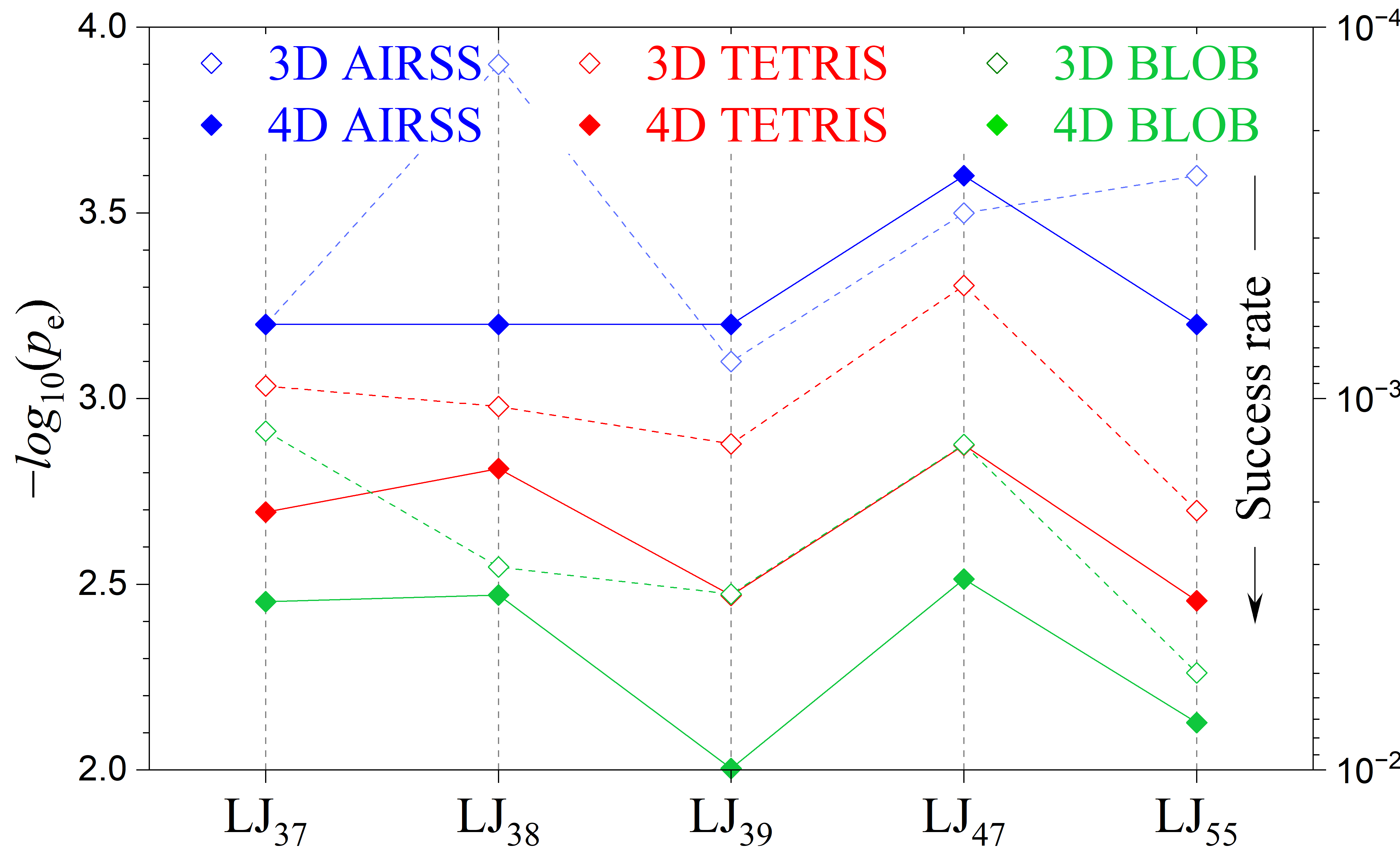}
    \caption{\label{fig-03} Success rates of finding global minima for LJ clusters using different structure generation protocols (BLOB versus TETRIS) and optimization methods (3D versus 4D). The AIRSS results were taken from Ref.~\cite{Pickard2019}. Each run contained between 10$^4$ and $2.5\times10^5$ structures that ensured the location of the ground state at least 100 times. Lower $-\log_{10}{p_e}$ values mean higher success rates.}
\end{figure}

\subsection{Lennard-Jones clusters}

Over a century, the LJ potential has served as a canonical model for systems with isotropic interactions and negligible many-body effects. Despite the simplicity of the functional form, the ground states of LJ clusters, established at least up to at $N = 150$ atoms~\cite{Wales1997, Leary1999, Leary2000}, exhibit surprising structural complexity and have become standard benchmarks for developing and tuning global optimization algorithms~\cite{Wales1997, Leary1999, Leary2000, DAVEN1996, Brüning2009, Rondina2013, Pickard2019, Tao2024}. 
 
We applied our implementations of the TETRIS and BLOB structure generation methods along with hyperdimensional optimization to five LJ clusters with specified point group symmetries: 37 ($C_1$), 38 ($O_{h}$), 39 ($C_{5v}$), 47 ($C_1$), and 55 ($I_{h}$). These sizes were selected due to their nontrivial geometries, such as LJ$_{38}$ known for its double-funnel energy landscape~\cite{doye1999}. Fig.~\ref{fig-03} shows the success rates of global minimum searches using various structure generation protocols and optimization strategies. To ensure statistical significance and direct comparison with the original GOSH study, we generated sufficient structure sets, up to $2.5 \times 10^5$ per size, to reach 100 ground state hits and picked the highest reported success rates obtained with AIRSS for packing fractions ranging between $0.1$ and $0.3$, as denser packing was determined to be not universally beneficial for locating ground states~\cite{Pickard2019}. In fact, the consistently high compressions achieved with our TETRIS generation scheme (Supplementary Fig.~3~\cite{SM}) were apparently detrimental for all considered sizes, as starting configurations constructed with the BLOB protocol converged to global minima $\sim 2.3$ times more frequently.

In contrast to the reported GOSH results, the addition of the extra coordinate through MAP0 improved the convergence across {\it all} considered sizes and seed generation strategies, effectively doubling the average probabilities of finding the LJ global minima. Unsurprisingly, the highest success rates, up to 1 in 100 tries, were observed for LJ$_{39}$ and LJ$_{55}$ with the most symmetric global minimum configurations. Histograms of local minima energies collected in Supplementary Fig.~4~\cite{SM} establish that the extra dimension helps LJ clusters transform into more favorable configurations not just near the bottom of the potential energy surface but across the full landscape. The presented findings for different sizes and structure initializations demonstrate that systematic gains in hyperspatial optimization of LJ clusters can be attained through relatively modest $\pm 1$ \AA\ maximum displacements of preconstructed 3D configurations into the extra dimension.

The net performance should be assessed by taking into account the increased computational cost of the introduced structure generation and optimization algorithms. According to our estimates, the BLOB and TETRIS operations take one to two orders of magnitude less time than typical local optimizations with the LJ potential. However, hyperspatial optimization can lead to significant slowdowns due to the overhead introduced by the extra spatial dimension and the increased number of energy and force evaluations required for convergence. For the LJ potential, the dimensional overhead added only about 8\%, while the average number of relaxation steps increased by a factor of $413/212\approx 1.9$, offsetting but not fully negating the benefit of 4D optimization. 

\begin{figure}[t!]
   \centering
\includegraphics[height=0.3\textwidth]{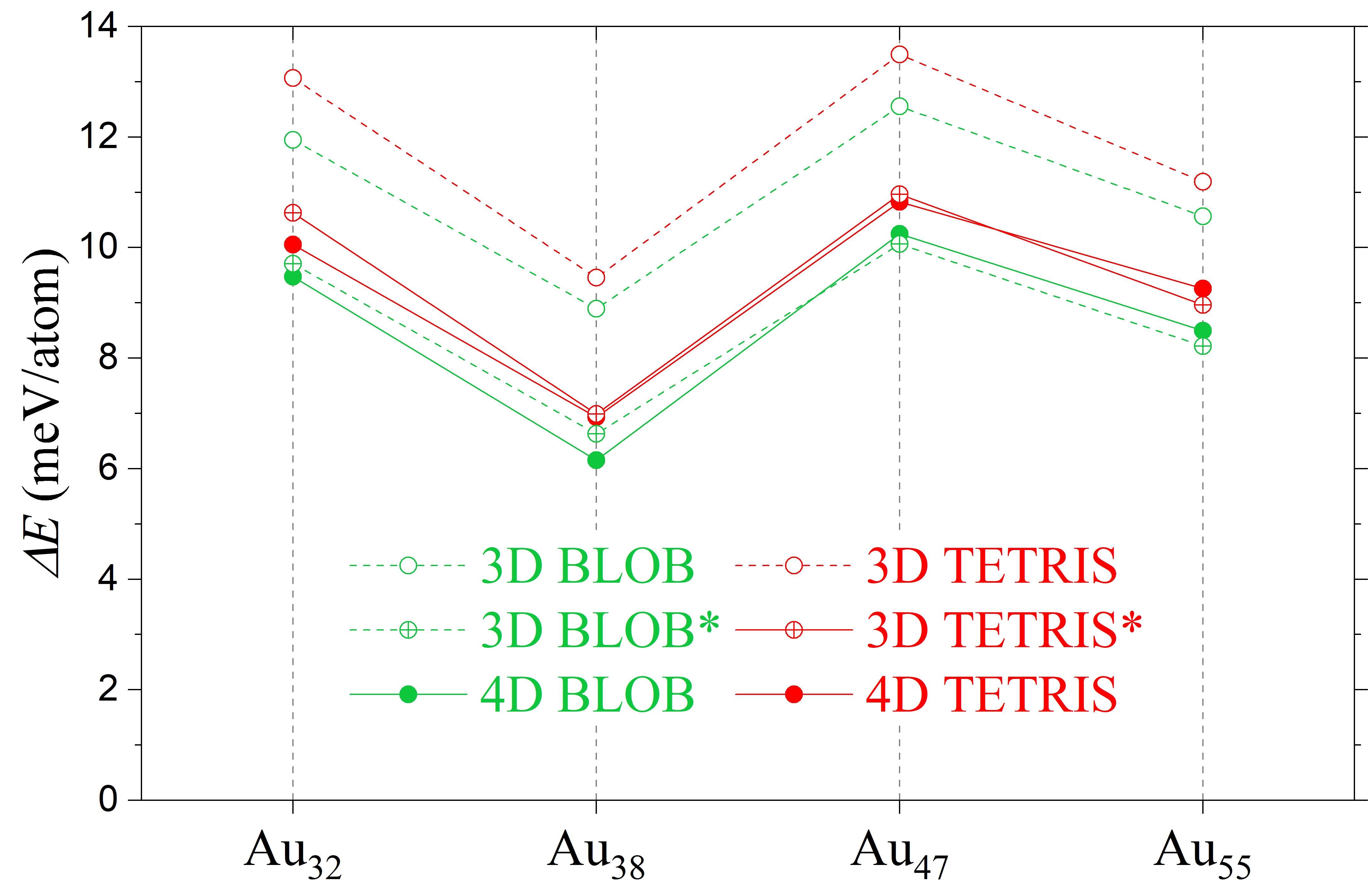}
  \caption{\label{fig-04} Average energies of the best 2,000 structures relative to the ground states for Au clusters modeled with NNPs. Each set generated with different protocols (random versus TETRIS) and optimized with different methods (3D versus 4D) consisted of 5,000 structures. The asterisk denotes CPU-time-adjusted average as explained in the text.}
\end{figure}

\subsection{Au clusters modeled with NNPs}
As the first testbed for benchmarking hyperdimensional optimization on a chemically realistic many-body system, we chose small Au clusters described by a NNP. Nanosized Au has a complex potential energy surface due to relativistic effects and strong d-orbital hybridization, which lead to low-symmetry ground states including planar, hollow, and amorphous morphologies~\cite{Garzon1996,Johansson2004,Wang2005,TARRAT2017}. Accurate structure prediction is particularly relevant for nanoparticles of this noble metal, as their rich catalytic properties are known to be highly sensitive to geometry~\cite{LOPEZ2004,Suchomel2018}. Extensive tests in our previous study established that the conventional Gupta potential or embedded atom model are inadequate for guiding {\it ab initio} structure searches, with Gupta-potential minima dispersing by nearly 30 meV/atom upon DFT relaxations~\cite{ak40}. In contrast, our NNP trained to 6.5 meV/atom accuracy enabled refinement of ground-state candidates across the $30 \leq N \leq 80$ range and helped identify a more stable amorphous configuration for Au$_{55}$.

Since NNPs are roughly two orders of magnitude more expensive than the LJ potential and their ground states are subject to particular parametrization, we adopted a different benchmark strategy for Au clusters. Rather than measuring success rates based on the successful identification of global minima, we generated 5,000 structures for each cluster size and method and calculated the average energy of the lowest 40\% (2,000) minima relative to the best NN-relaxed structure obtained in this dataset. This metric reflects the sampling efficiency of each method while remaining robust to uncertainty in the true ground-state configurations. For example, the combined $2\times 10^4$ tries for each size summarized in Fig.~\ref{fig-04}, reproduced the putative NN ground state for Au$_{47}$, converged to suboptimal minima for Au$_{32}$ and Au$_{55}$, and identified a more favorable configuration, by $\sim 1.7$ meV/atom, for Au$_{38}$ compared to the prior multi-tribe evolutionary searches with $2.5\times 10^4$ local optimizations~\cite{ak40}.

Across all tested cluster sizes and initialization strategies, 4D optimization yielded average energy reductions of approximately 2.5 meV/atom compared to 3D relaxation. However, this gain is counterbalanced by an average 1.63-fold increase in the number of iterations leading to an average 1.78-fold increase in total wall time per relaxation. To account for this difference in computational cost, we reevaluated the effective sampling performance by proportionally reducing the size of the low-energy pool used for averaging. Namely, we adjusted the 3D data by computing the average over the lowest 2000/1.78 structures instead of 2,000. When this correction is applied, the apparent advantage of 4D optimization largely vanishes, indicating that the observed improvements stem from extended local relaxation rather than more efficient exploration.

\begin{figure}[!t]
   \centering
\includegraphics[width=0.445\textwidth]{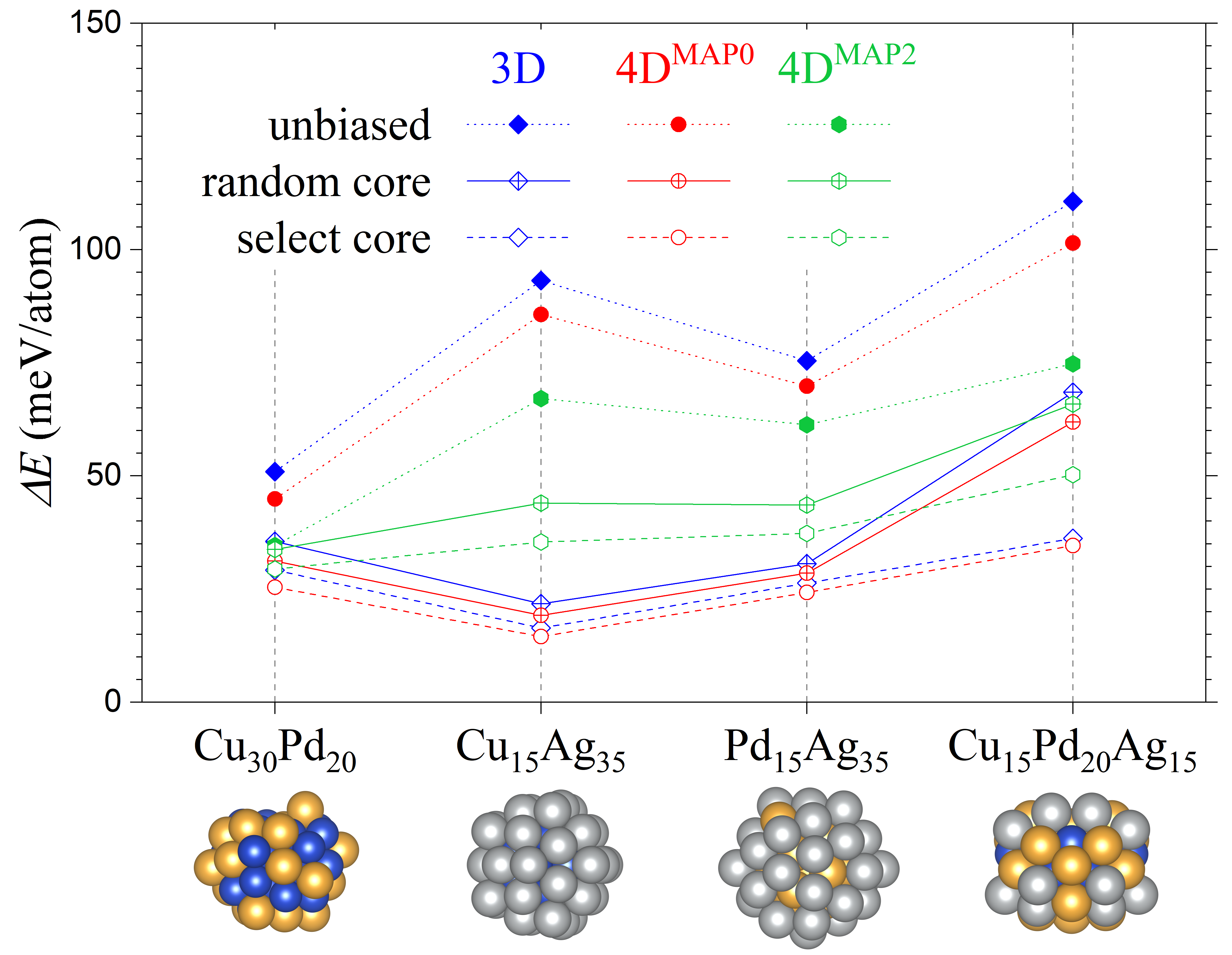}
  \caption{\label{fig-05} Average NN energies of the best 2,000 out of 5,000 structures relative to the ground states for 50-atom clusters found in Ref.~\cite{ak38} and displayed with blue (Cu), yellow (Pd), and silver (Ag) spheres. The solid, crossed, and open symbols correspond to clusters generated with TETRIS using different species sequences as explained in the text. Visual representations of atomic structures were created with VESTA~\cite{vesta}.}
\end{figure}

\subsection{Cu-Pd-Ag clusters modeled with NNPs}

Nanoalloys display a broader spectrum of tunable catalytic, optical, and magnetic properties due to their diverse composition-dependent core-shell, Janus, and mixed-species arrangements. To evaluate the performance of the hyperdimensional and TETRIS methods in identifying the most favorable configurations of multicomponent nanoparticles, we focused on 50-atom binary and ternary Cu-Pd-Ag clusters. Our previous investigation of the system with the evolutionary algorithm coupled with NNPs revealed a highly frustrated energy landscape across the composition range, making it a challenging benchmark for global optimization~\cite{ak38}. The reported most favorable Cu-Pd-Ag configurations have core-shell morphologies, with the larger metals preferentially residing on the surface (see Fig.~\ref{fig-05}).

Fig.~\ref{fig-05} illustrates how the choice of local optimization strategy and initial population seeding influences search efficiency. All starting structures, 5,000 for each data point, were generated using the TETRIS approach, but species assignment during cluster assembly followed three different schemes. In the unbiased case, species were assigned in a completely random order. In the random core variant, atoms were grouped by species, but the sequence of species was randomized. In the select core protocol, species were ordered by atomic size, from smallest to largest. We chose to average the energies over the best 2,000 minima, i. e., 40\% of each set, to clarify the performance of the random core generation for binary or ternary nanoalloys which have $1/(2!)=50$\% or $1/(3!)\approx 17$\% chances of having the optimal core-shell sequences.

In unbiased searches, 4D optimizations consistently produced lower-energy structures than standard 3D relaxations across all tested nanoalloys. The average energies above the global minima decreased by $\sim 10$\% with MAP0 and $\sim30$\% for MAP2, indicating that the added dimension, with larger average displacements in the latter case, allows the system to escape inefficient local geometries during relaxation. As in the Au case, the additional computational overhead, now measured at 2.24 for MAP0 and 3.25 for MAP2, diminishes these gains but the full factors behind this improvement become clearer when considered alongside the results with the other two structure initialization protocols. Growing clusters by grouping incoming atoms by species, either randomly or selectively, dramatically improved the search outcomes for both 3D and 4D optimizations. The moderate randomization of the extra dimensional coordinate up to $\pm 1$ \AA\ in MAP0 continued to provide a $\sim 10$\% improvement, while the cluster ‘spherization’ in all dimensions in MAP2 likely promoted a certain degree of species mixing and led to less favorable minima than in the standard 3D case. The random and select core schemes yielded fairly similar results for binary clusters, which is consistent with the 50\% probability of obtaining the correct species sequence when two species are randomly ordered. In contrast, the difference was more pronounced for ternary clusters, where the chance of realizing the correct sequence under random grouping drops to about 17\%. Compared to our previous evolutionary searches with $5\times10^4$ local optimizations per nanoalloy starting with unbiased TETRIS configurations~\cite{ak38}, the present local relaxations totaling $4.5\times10^4$ tries yielded comparable best minima for binary clusters, 0.1 meV/atom higher for Cu$_{30}$Pd$_{20}$, the same for Cu$_{15}$Ag$_{35}$, and 0.3 meV/atom lower for Pd$_{15}$Ag$_{35}$, but a significantly less optimal minimum, by 7.5 meV/atom, for Cu$_{15}$Pd$_{20}$Ag$_{15}$.

These trends suggest that the key limitation of standard 3D relaxation lies not in geometric constraints, but in its inability to reassign species to more favorable environments. Just as in the optimization of binary chains~\cite{Pickard2019}, GOSH circumvents this by effectively enabling atom reordering during relaxation, but the same benefit in the more practical case of multicomponent nanoparticles can be attained more directly through incorporation of chemical knowledge in the initial population. These observations align with alternative efforts in the field to address compositional frustration in multicomponent clusters. Several methods have been developed to enable or approximate species swaps during optimization, including identity-exchange Monte Carlo steps in basin-hopping~\cite{Rondina2013}, symmetry-constrained genetic algorithms~\cite{Han2022}, and multi-agent search protocols with specialized roles such as the Flying-Landing-Hiking framework~\cite{Rapetti2023}.

\subsection{Li-Sn, Pd-Sn, and Ag-Sn modeled with NNPs}

To test the performance of global optimization in periodic systems, we selected M-Sn intermetallics known to display a wide range of morphologies. As a group XIV element capable of covalent bonding, Sn readily forms directional extended frameworks and cage-like motifs, especially when alloyed with electropositive metals~\cite{Nylen2004, Engelkemier2013,ak31,ak47,ak51}. For example, Sn-rich alloys can be represented with particular sequences of Kepler nets, while M-rich intermetallics often appear as simple decorations of common metallic lattices with isolated or only weakly connected Sn atoms. Our recent NNP-guided exploration of M-Sn binaries (M = Li, Na, Ca, Cu, Pd, and Ag) led to the identification of over 30 new phases stable under different $(T,P)$ conditions~\cite{ak47,ak51}. They include tI36-PdSn$_2$ with an A$^\circ$B$^\circ$A$^-$A$^\circ$B$^\circ$A$^+$ stacking of Pd atoms between Sn layers with rotated squares and hR75-Li$_{19}$Sn$_6$ with a [345454] stacking of layered units defining the bcc decoration.

We focused on Li$_7$Sn$_9$, Li$_9$Sn$_6$, Pd$_4$Sn$_{12}$, Pd$_{14}$Sn$_2$, and Ag$_{14}$Sn$_4$ compositions determined in our prior evolutionary searches to have complex ground states with sizes between 15 and 18 atoms per unit cell. To gauge the capability of the TETRIS method, we tried only broadly applicable structural fragments, such as M–M, M–Sn, and Sn–Sn dimers or bcc blocks, avoiding overly material-dependent structural motifs that could bias the searches toward previously known solutions. Our tests (not shown) demonstrated little advantage of block seeding over random structure generation, and we chose the latter in our investigation of the hyperspatial optimization.

\begin{figure}[t!]
   \centering
\includegraphics[width=0.445\textwidth]{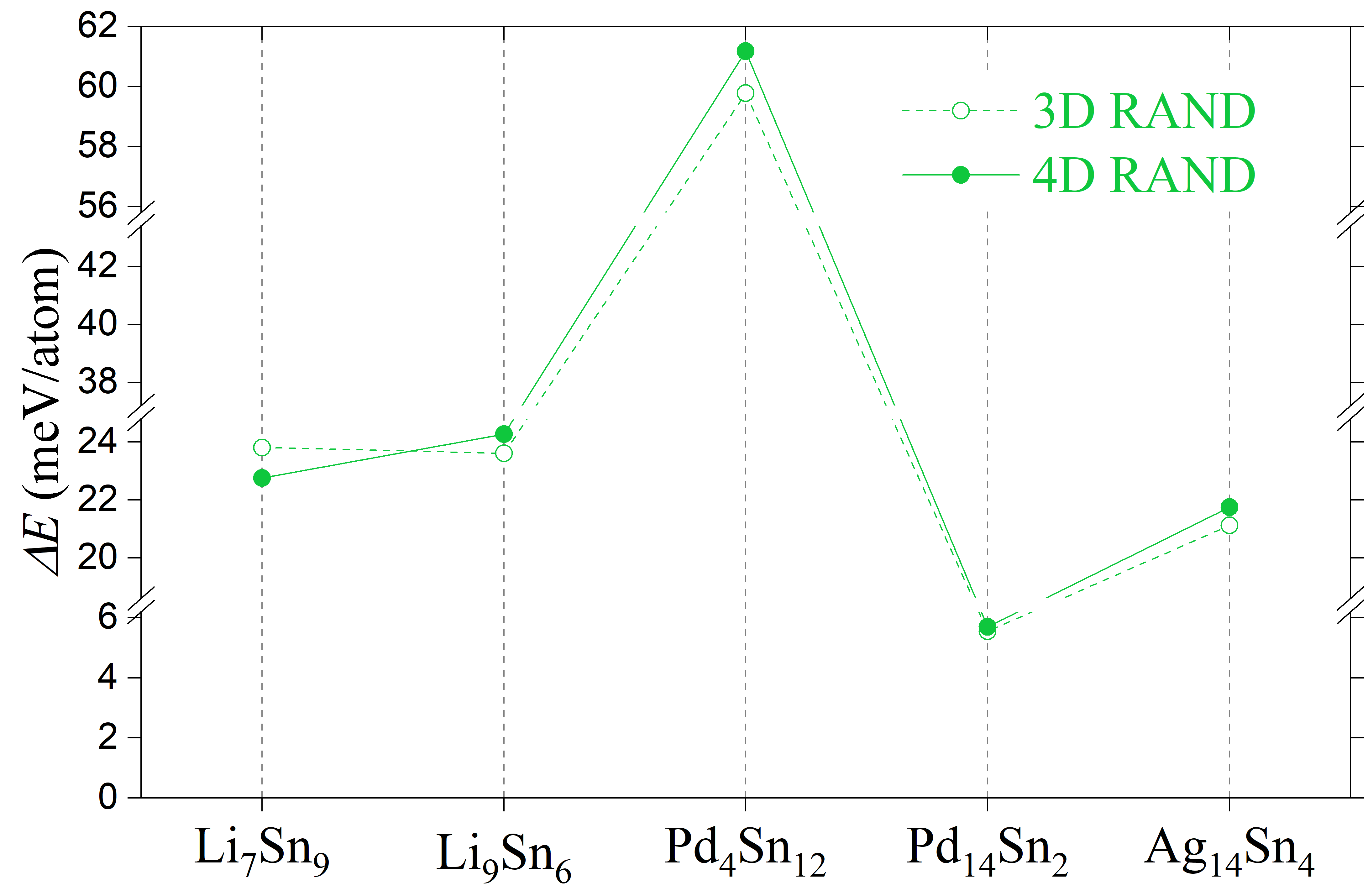}
  \caption{\label{fig-06} Average energies of the best 2,000 structures relative to the ground states for Li-Sn, Pd-Sn, and Ag-Sn crystals modeled with NNPs. Each set optimized with different methods (3D and 4D) consisted of 5,000 structures.}
\end{figure}

Figure~\ref{fig-06} compares the average energies of the 2,000 lowest structures out of 5,000 relative to the ground state, evaluated by neural network potentials. Unlike in nanoalloys, 4D optimization did not yield systematic improvements in periodic crystals. Both 3D and 4D relaxations sampled similar energy distributions across the lowest 40\% of minima and delivered comparable success in locating the lowest-energy states identified in earlier evolutionary runs. For Li$_7$Sn$_9$ and Pd$_{14}$Sn$_2$, both methods repeatedly recovered the ground state. For Pd$_4$Sn$_{12}$, only 4D found the ground state, with 3D reaching a structure 2 meV/atom higher. For Ag$_{14}$Sn$_4$, the ground state was found only by 3D, with the closest 4D result 0.5 meV/atom less stable. The lack of significant advantage observed for 4D relaxation in these periodic systems may result from the indirect constraints imposed by lattice vectors, which evolve primarily in response to effective three-dimensional interatomic distances modulated by the additional hyperspatial coordinates.

\subsection{LiB$_2$C$_2$ and CaB$_6$ modeled with DFT}
The usefulness of the TETRIS adaptation to periodic systems can be illustrated with {\it ab initio} searches for ground states in complex covalent materials, such as metal borides and borocarbides known to display a variety of 2D and 3D frameworks~\cite{Kouvetakis1986,ROGL2014}. To collect sufficient data for benchmarking different protocols at the DFT level, we generated 1,000 or 2,000 medium-sized configurations and assessed each variant’s performance by calculating the average energy of the best 2\% of locally optimized candidates. The dataset sizes were chosen to approximate fractions of random motifs injected into population during typical evolutionary runs totaling 4,000-6,000 local optimizations~\cite{ak23,ak41}. Our recent screening of layered metal borocarbides for conventional high-$T_{\textrm c}$ superconductors synthesizable under ambient pressure revealed that high levels of hole doping destabilize the honeycomb layer morphology in M$_x$BC (M = Li, Na, or Mg) essential for strong quasi-2D electron-phonon coupling~\cite{ak52,ak54}. At the Li$_{1/2}$BC composition, our evolutionary searches identified an mS20 ($Cm$) structure with a 3D framework $\sim 70$ meV/atom below the best oI10 ($Imm2$) layered counterpart. 

\begin{figure}[t!]
   \centering
\includegraphics[width=0.45\textwidth]{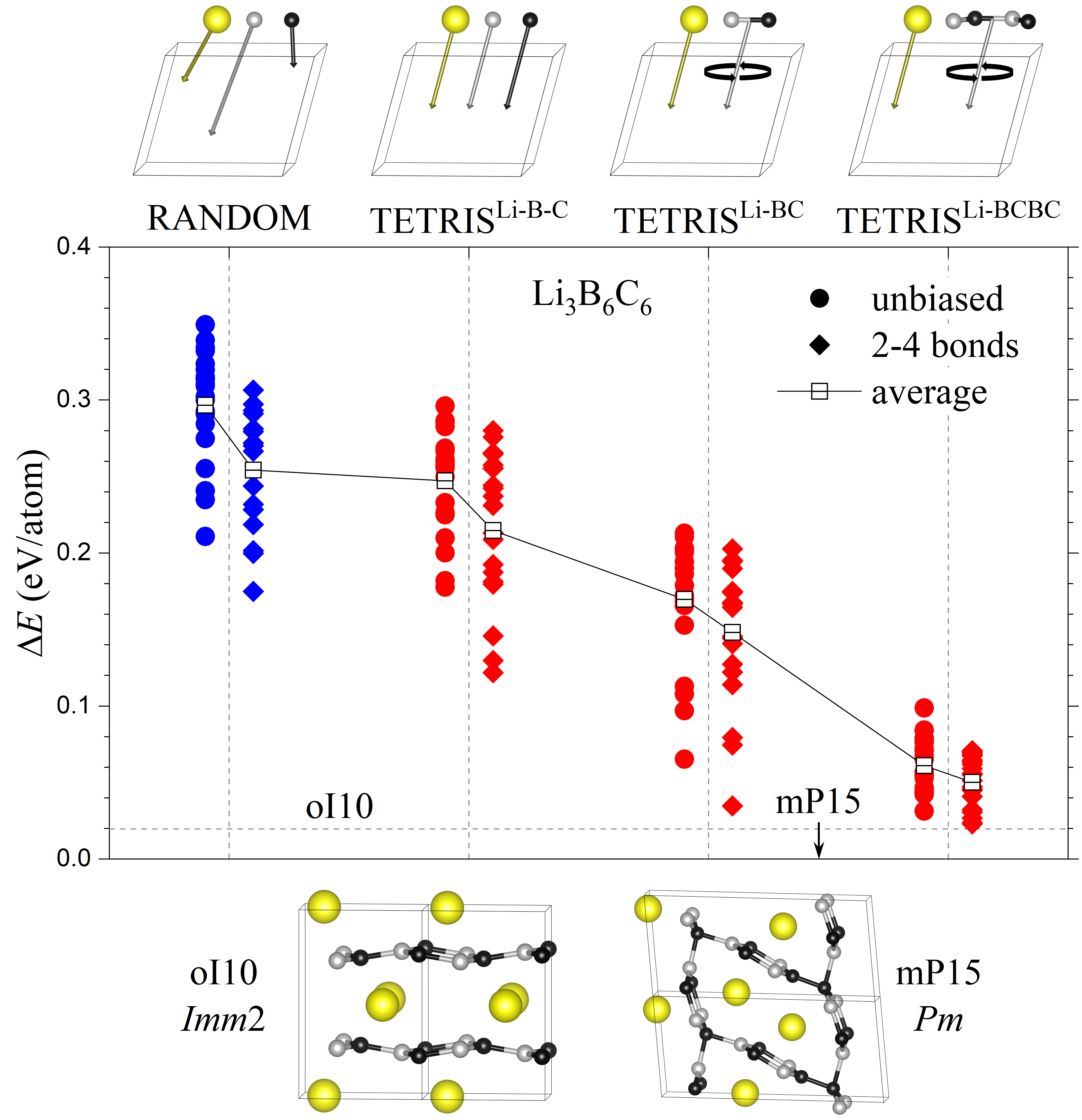}
    \caption{\label{fig-07} Performance of random and TETRIS structure generation methods in the search for most favorable Li$_3$B$_6$C$_6$ configurations at the DFT level. The number and types of TETRIS blocks are indicated at the top. The sets denoted with diamonds were generated with an additional constraint for B and C atoms to have 2-4 neighbors within 1.9~\AA. The energy is referenced to the most favorable mP15-Li$_3$B$_6$C$_6$ phase with 3 formula units found in our previous study~\cite{ak52}, with the dashed line corresponding the lowest-energy layered configuration.}
\end{figure}

We performed our present tests for Li$_3$B$_6$C$_6$ and referenced the resulting $8\times 1,000$ structures to the mS15 ($Pm$) polymorph with 3D morphology determined previously to have the lowest energy among 15-atom unit cells. Fig.~\ref{fig-07} shows the effectiveness of generating low-energy motifs as a function of the introduced bias. For each protocol, we checked whether the additional condition on the number of nearest neighbors, 3-4 within 1.9 \AA\ around B/C atoms expected to be in the sp$^2$ or sp$^3$ environments, leads to better starting configurations. The constraint indeed consistently lowered the average energy, with the largest improvement of $\sim 40$ meV/atom noted for the random scheme. The three TETRIS sets with different numbers of atomic blocks may help find a desired balance for bias in structure generation. A simple switch from a random placement of atoms within a unit cell to the TETRIS packing of individual atoms resulted in a noticeable drop of the average energy by $\sim 50$ meV/atom. The introduction of B and C atoms in pairs to promote a natural alternating order in heteronuclear frameworks further lowered the average energy by $\sim 75$ meV/atom. The larger four-atom preassembled units link into honeycomb layers far more efficiently while not prohibiting interlayer bridging, as we observed low-energy configurations with sp$^3$-bonded C atoms.

\begin{figure}[t!]
   \centering
\includegraphics[width=0.45\textwidth]{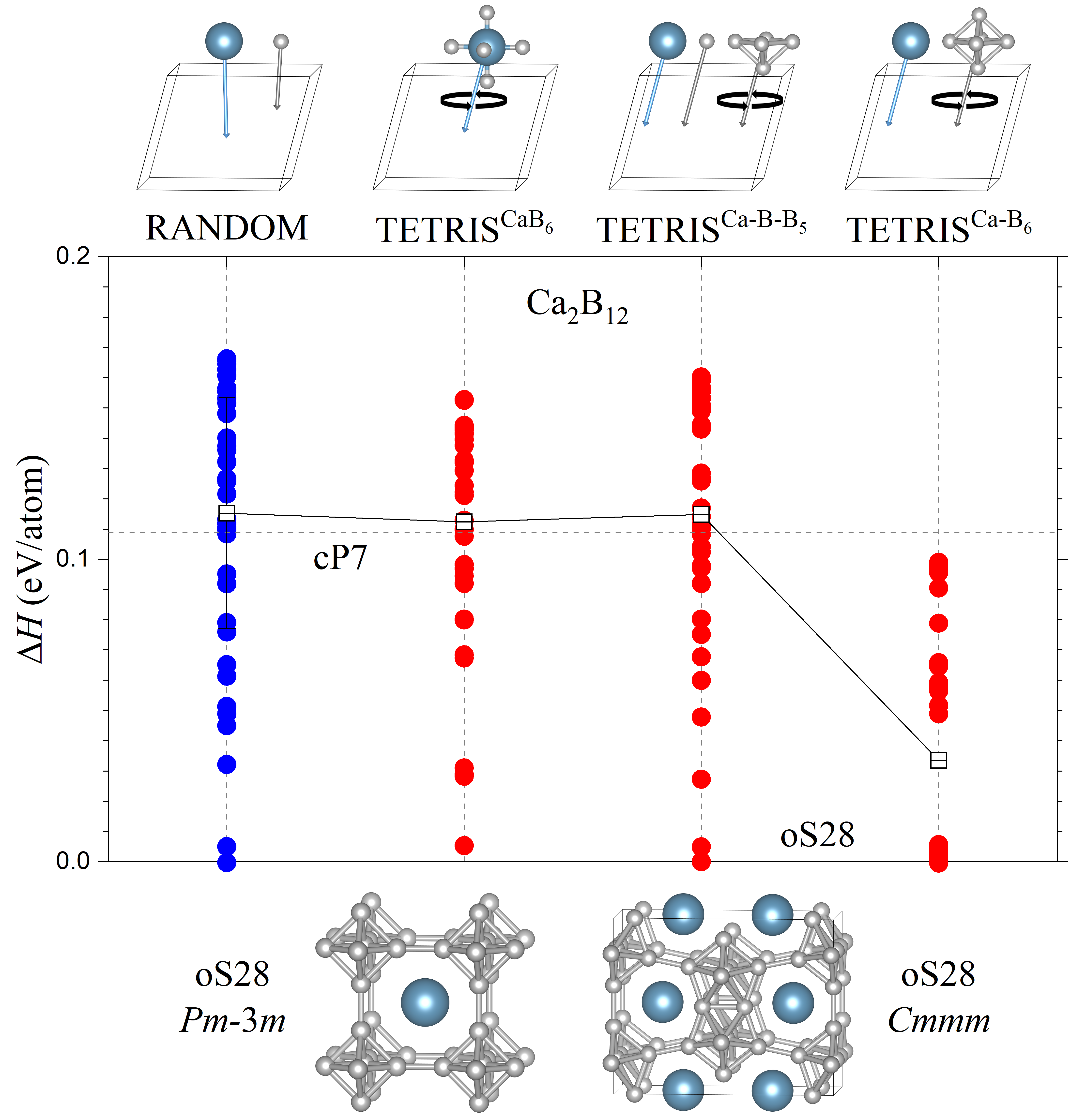}
    \caption{\label{fig-08} Performance of random and TETRIS structure generation methods in the search for most favorable Ca$_2$B$_{12}$ configurations at 50 GPa at the DFT level. The number and types of TETRIS blocks are indicated at the top. The enthalpy is referenced to the most favorable oS28-Ca$_2$B$_{12}$ phase found in our previous study~\cite{ak23}, with the dashed line corresponding to the cP7 ground state at ambient pressure.}
\end{figure}

Investigation of CaB$_6$ under pressure provides information on the methods’ performance for non-layered configurations. Our previous studies revealed exceptional complexity of pressure-induced structural transformations in this compound, which starts as a simple cP7 at ambient pressure, becomes dynamically unstable above 23 GPa, and displays a variety of more stable exotic 3D frameworks above 13 GPa~\cite{ak23,ak24}. The concurrent synthesis work indicated the formation of a new phase above 31 GPa but its structure could not be established from the collected XRD data and was solved with evolutionary searches without any structural input~\cite{ak23}. Our benchmark tests demonstrated that the construction of the 28-atom tI56 ground state could not be accomplished with unbiased random searches in 10,000 tries and required either randomization of the cP7 supercells or use of the evolutionary algorithm. 

Here, we considered smaller 14-atom Ca$_2$B$_{12}$ unit cells at 50 GPa, determined to have oS28 ($Pm$-$3m$) as the lowest-enthalpy configuration with up to two formula units~\cite{ak23}, and allowed an additional 3D random rotation of the TETRIS blocks with quaternions in their starting position. Fig.~\ref{fig-08} displays the average enthalpy of the best 40 structures and shows that 2,000 tries were sufficient to locate oS28 with each protocol. In fact, the TETRIS algorithm offered no improvement over the random method when we used blocks with one-three B atoms, a preassembled unit with a uniform distribution of six B atoms around Ca, or a five-atom fragment of the B octahedron (only the last two sets are shown). An appreciable enthalpy drop of nearly 80 meV/atom was observed only when full octahedron units were used. 

We would like to stress that we view the TETRIS generation not as a stand-alone optimization method but rather as a protocol for seeding global structure searches with plausible configurations. The presented results of {\it ab initio} searches demonstrate that steering global exploration toward layered configurations can be accomplished naturally with minimal bias, such as heteronuclear dimers. Construction of complex 3D frameworks may require the use of more specific building blocks favorable for this material’s class but the TETRIS packing is not constrained to supercells of known prototypes to generate periodic configurations. 

\section{Summary}
\label{sec:conclusions}

We present a systematic extension and examination of two complementary strategies for improving global minima searches in nanoscale and crystalline systems: structure generation with physically biased seeding and hyperspatial optimization. To enable treatment of chemically realistic systems with a variety of structure initialization and minimization schemes, we expanded Behler-Parrinello-type NNPs to extra dimensions and reformulated several elements of the GOSH algorithm. Specifically, we introduced two coordinate mapping schemes, MAP0 and MAP2, that displace physically meaningful configurations constructed in the normal 3D space into extra dimensions using distinct extension strategies. MAP2 mimics the original GOSH approach by generating randomized displacements that effectively treat all spatial directions equally, producing near-hyperspherical distributions. In contrast, MAP0 introduces moderate, uncorrelated shifts of up to $\pm 1$ \AA\ into the extra dimensions. Additionally, we adjusted the protocol for changing the spring constant associated with extraspatial coordinates to stepwise reductions during local optimizations, enabling the use of advanced minimization algorithms, such as BFGS2.

Our benchmark tests showed that both the modified mapping scheme and enhanced relaxation protocol were essential for attaining consistent gains in complex systems. In multicomponent Cu-Pd-Ag clusters initialized with unbiased species ordering, 4D optimization with the conservative MAP0 approach reduced the average energy above the global minimum by $\sim10$\%, while the more aggressive MAP2 scheme yielded up to $\sim30$\% reductions, highlighting the benefit of using extra dimensions to facilitate atom swaps. However, a chemically informed TETRIS initialization, which assembles clusters with species grouped by atomic size to favor core-shell formation, led to significantly better convergence even with standard 3D optimization, and performed even more effectively when paired with MAP0. This balance between geometric flexibility and controlled species placement was especially important for ternary systems, where the probability of randomly assembling optimal core-shell configurations is inherently low. For bulk alloys, these factors had little impact, as neither hyperspatial optimization nor species ordering strategies produced measurable gains. In contrast, TETRIS-based seeding, by assembling unit cells from chemically sensible atomic blocks, substantially improved convergence in covalently bonded systems modeled with DFT.

This study provides the first evidence that hyperspatial optimization, previously applied to simple potential energy surfaces~\cite{Faken1999,Bayden2004,Pickard2019}, can yield benefits in systems modeled with sophisticated many-body potentials. By independently varying the coordinate mapping and initialization strategies, we were able to disentangle the contributions of geometric relaxation flexibility and species ordering, a distinction particularly important for nanoalloys. These insights offer a path toward improving hyperspatial optimization across a broader range of materials systems.

\begin{acknowledgments} 
The authors acknowledge support from the National Science Foundation (NSF) (Award No. DMR-2320073). This work used the Expanse system at the San Diego Supercomputer Center via allocation TG-DMR180071. Expanse is supported by the Extreme Science and Engineering Discovery Environment (XSEDE) program~\cite{XSEDE} through NSF Award No. ACI-1548562.

\end{acknowledgments}

\newpage
\newpage

\bibliography{refs} 
\newpage

\end{document}